\documentclass[prl,showpacs,amsmath,assymb,eufrak]{revtex4}
\usepackage{epsfig,amsfonts,amsbsy}

\newcommand{\sqbk}[1]{\left[#1\right]}
\newcommand{\bsqbk}[1]{\left[#1\right]}

\newcommand{\sbkt}[1]{\langle#1\rangle}
\newcommand{\bbkt}[1]{\bigl\langle#1\bigr\rangle}

\newcommand{\WT}{\mathcal{T}}
\newcommand{\calD}{\mathcal{D}}

\newcommand{\pathG}{\hat{\Gamma}}

\newcommand{\pathGi}{\pathG_\mathrm{i}}
\newcommand{\pathGm}{\pathG_\mathrm{m}}
\newcommand{\pathGmd}{\pathG_\mathrm{m}^\dagger}
\newcommand{\pathGf}{\pathG_\mathrm{f}}
\newcommand{\patha}{\hat{\nu}}
\newcommand{\pathad}{\hat{\nu}^\dagger}

\newcommand{\Jv}{J_\mathrm{viol}}
\newcommand{\Qt}{Q_\mathrm{t}}
\newcommand{\rhost}{\rho^\mathrm{st}}

\newcommand{\mbeta}{\bar{\beta}}
\newcommand{\dbeta}{\Di\beta}
\newcommand{\oep}{O(\epsilon^3)}
\newcommand{\oet}{O(\epsilon^2)}

\newcommand{\Di}{\mathit{\Delta}}

\newcommand{\pa}{\nu}
\newcommand{\tr}{\tau_\mathrm{r}}
\newcommand{\tl}{\tau_\ell}
\newcommand{\ts}{\tau_\mathrm{s}}
\newcommand{\tob}{\tau_\mathrm{o}}

\newcommand{\Gi}{\Gamma_\mathrm{i}}

\newcommand{\intt}{\int_{-\tl}^{\tl} dt}
\newcommand{\intti}{\int_{-\tl}^{-\ts} dt}
\newcommand{\inttf}{\int_{\ts}^{\tl} dt}
\newcommand{\inttm}{\int_{-\ts}^{\ts} dt}
\newcommand{\intto}{\int_{-\tob}^{\tob} dt}
\newcommand{\ste}{\mathrm{st}}

\begin{document}
\title{
Work Relation and the Second Law of Thermodynamics in Nonequilibrium Steady States
}
\author{Naoko Nakagawa}
\affiliation {
College of Science,  Ibaraki University, Mito, Ibaraki 310-8512, Japan
}
\date{January 15, 2012}

\begin{abstract}
We extend Jarzynski's work relation and the second law of thermodynamics 
to a heat conducting system which is operated by an external agent.
These extensions contain a new nonequilibrium contribution
expressed as the violation of the (linear) response relation
caused by the operation.
We find that a natural extension of the minimum work principle involves 
information about the time-reversed operation, and is far from straightforward.
Our work relation may be tested experimentally especially when 
the temperature gradient is small.
\end{abstract}


\maketitle

\section{Introduction}

Thermodynamics is a universal framework for macroscopic systems in equilibrium.
The second law, which is at the heart of thermodynamics, gives strict limitations
to macroscopic operations and provides fundamental concepts such as irreversibility
and minimum work.
The idea of minimum work leads to the useful Gibbs relation, which represents work
associated with a thermodynamic operation as the difference in the free energy.
To develop similar useful thermodynamics for nonequilibrium systems is a
fascinating challenge.
In \cite{Oono-Paniconi,Sasa-Tasaki,Ruelle,Hatano-Sasa,KN,KNST,KNST-nl}, attempt has been made to construct operational thermodynamics for nonequilibrium steady states (NESS).
A central idea was to replace the ``bare heat'' in NESS by its ``renormalized'' counterpart called excess heat.

Recently there has been a considerable progress in nonequilibrium physics
which in particular led to the fluctuation theorem \cite{symmetry2}
and the Jarzynski equality \cite{Jarzynski,Crooks}.
The former gives an exact equality for the entropy production in NESS,
which is connected to response relations.
The latter provides an exact relation between operational work and the free energy 
not only in quasi-static but also in general operations in equilibrium.
It is also directly connected to the second law of thermodynamics.

In the present paper, we focus on mechanical work associated with an external operation
in NESS realized in classical heat conducting systems.
We derive a very natural extension \eqref{e:Jarzynski} of the Jarzynski equality to NESS,
which may be tested experimentally.
The equality contains a new nonequilibrium contribution 
besides the usual contribution of the mechanical work and the free energy.
The new contribution is expressed as the violation of the linear response relation
caused by the external operation, which is represented by $\Jv(t)$  in \eqref{e:LRviolation}.
The extended equality straightforwardly implies the Gibbs relation \eqref{e:Gibbs} for the quasi-static limit and the second law \eqref{e:2ndLaw}.
The derivation of the results are essentially straightforward and is based on the detailed fluctuation theorem (also known as the microscopic reversibility or the local detailed balance condition).
We hope that these findings become crucial steps in the understanding and construction of thermodynamics for NESS.


\begin{figure}[t]
\begin{center}
\includegraphics[scale=0.4]{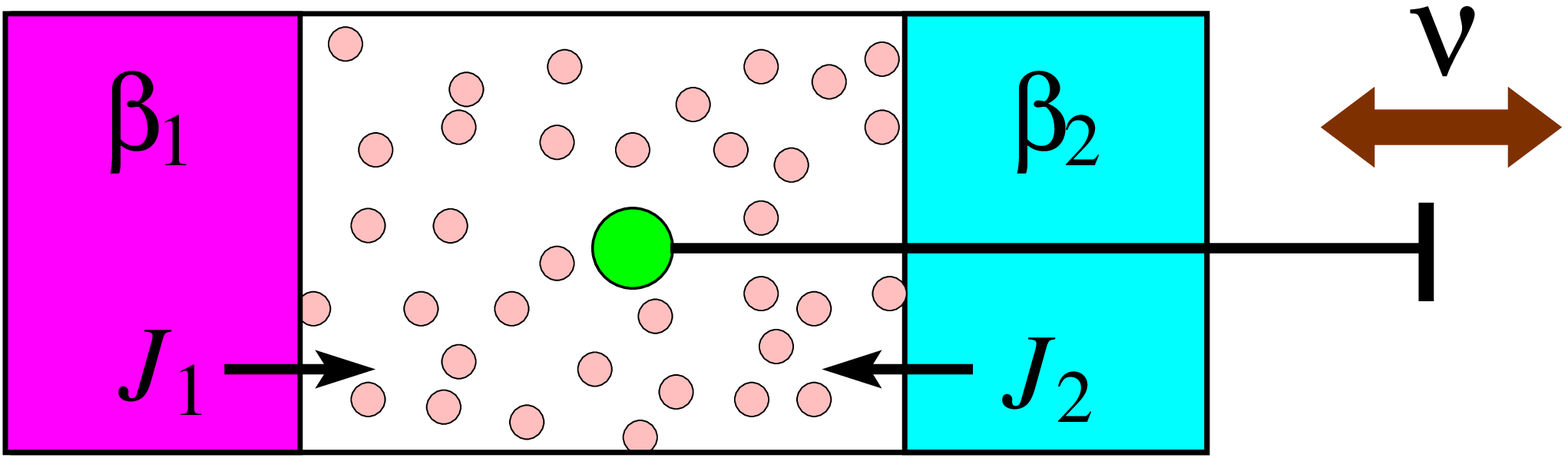}
\end{center}
\caption{(Color online) An example of the system, where the position $\nu$ of a particle (green (light gray))
can be controlled externally.
$J_1$ and $J_2$ are the heat currents from the heat baths to the system.}
\label{fig:fig1}
\end{figure}

\section{Setup}

Our theory can be developed in various nonequilibrium settings of classical stochastic systems.
For simplicity we here focus 
on heat conduction, and consider a system which is attached to two heat baths 
with inverse temperatures $\beta_1$ and $\beta_2$ and has  
controllable parameters $\nu$.
An example is a system of $N$ particles in a container 
in which the position $\nu$ of one of the particles 
is controlled by the external agent (see Fig.~1).
The inverse temperatures $\beta_1$ and $\beta_2$ are fixed throughout,
and are often omitted.
We define 
\begin{equation}
\mbeta:=\frac{\beta_1+\beta_2}{2}, \qquad \dbeta:=\beta_1-\beta_2.
\end{equation}

The coordinates of $N$ particles are collectively denoted as 
$\Gamma=(\mathbf{r}_1,\ldots,\mathbf{r}_N;\mathbf{p}_1,\ldots,\mathbf{p}_N)$,
and its time-reversal as
$\Gamma^*
=(\mathbf{r}_1,\ldots,\mathbf{r}_N;-\mathbf{p}_1,\ldots,-\mathbf{p}_N)$.
The time evolution of the system is governed by deterministic dynamics according to
the Hamiltonian $H_\pa(\Gamma)$ and
stochastic Markovian dynamics due to coupling to the two external heat baths.
We impose time-reversal symmetry $H_\pa(\Gamma)=H_\pa(\Gamma^*)$.
When discussing time evolution of $\Gamma$, we denote by $\Gamma(t)$ 
its value at time $t$,
and by $\pathG=(\Gamma(t))_{t\in[-\tl,\tl]}$ the path
in the whole time interval $[-\tl,\tl]$.
The heat baths may be realized in standard manners such as ``thermal walls'' 
(see the subsection below)
or the Langevin noise near the walls.
The only (and the essential) requirement is that the detailed fluctuation theorem (see \eqref{e:symmetry} below) is valid.
By $J_k(\pathG; t)$, we denote the heat current 
that flows from the $k$-th bath to the system
at time $t$ in the path $\pathG=(\Gamma(t))_{t\in[-\tl,\tl]}$
\cite{e:J}.
We write 
\begin{equation}
J(\pathG;t)=\frac{J_1(\pathG;t)-J_2(\pathG;t)}{2},
\end{equation}
 which is
the heat current from the first to the second heat bath.

\begin{figure}[b]
\begin{center}
\includegraphics[scale=0.4]{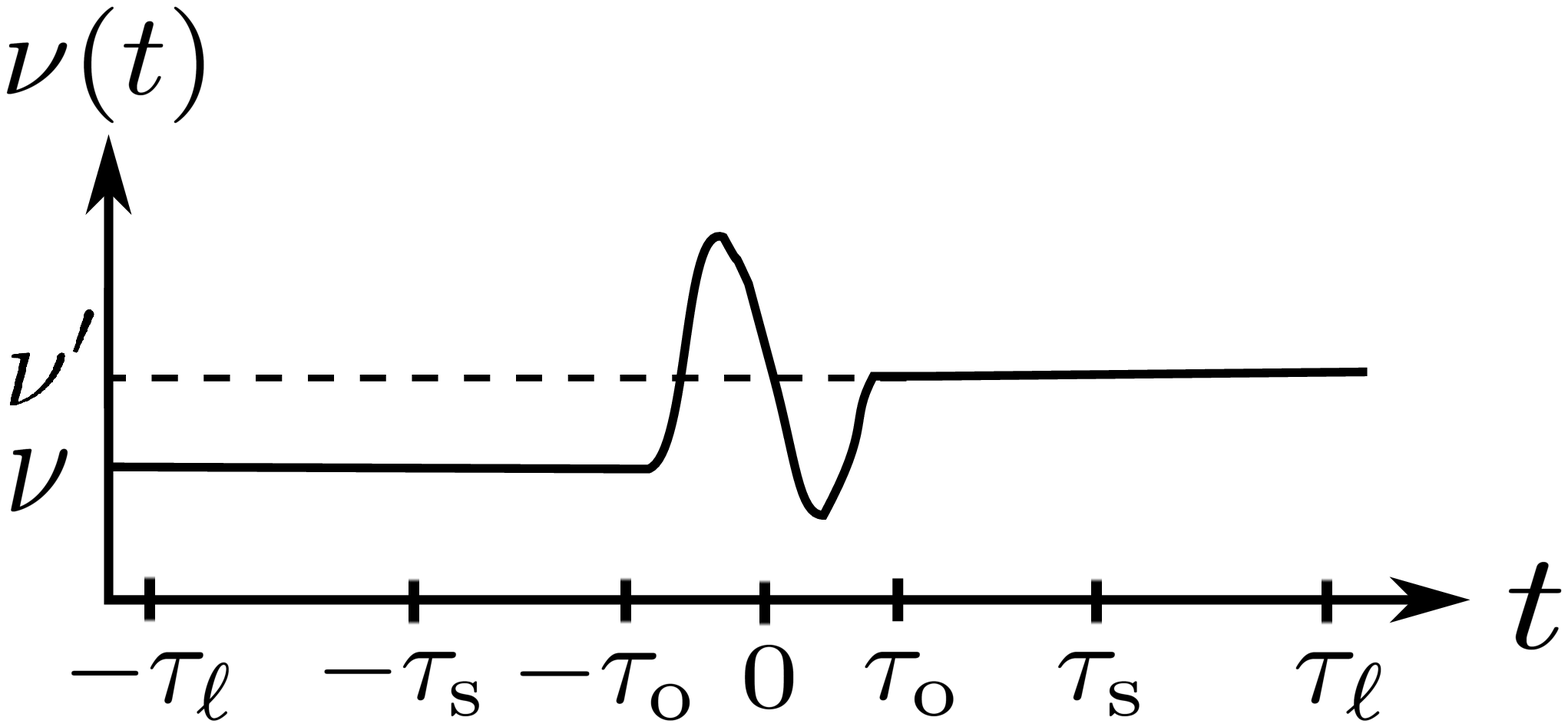}
\end{center}
\caption{A sketch of the protocol $\patha$.
We assume $\tl-\ts\gg \tr$ and $\ts-\tob\gg\tr$ where $\tr$ is the relaxation time of the system.
}
\label{fig:fig2}
\end{figure}

We shall assume that the system settles to a unique NESS 
when it evolves for a sufficiently long time with fixed $\nu$.
For later convenience we shall choose and fix three time scales $0<\tob<\ts<\tl$
such that $\tl-\ts\gg\tr$ and $\ts-\tob\gg\tr$
where $\tr$ is the relaxation time of the system.
See Fig.~2.
We suppose that an external agent performs an operation to the system 
by changing the parameters $\nu$ according to a prefixed protocol.
A protocol is specified by a function $\nu(t)$ of $t\in[-\tl,\tl]$.
In order to study transitions between NESS, 
we assume that $\nu(t)$ varies only for $t\in[-\tob,\tob]$, 
so that $\nu(t)=\nu$ for $t\in [-\tl,-\tob]$ and $\nu(t)=\nu'$ for $t\in [\tob,\tl]$.
We denote by $\patha=(\nu(t))_{t\in[-\tl,\tl]}$ 
the whole protocol,
by $\pathad=(\nu(-t))_{t\in[-\tl,\tl]}$ the time-reversal of $\patha$, 
and by $(\nu)$ the protocol in which the parameters are kept constant at $\nu$.
During the operation, the external agent performs mechanical work
\begin{equation}
W(\pathG)=\intto~ \left.\frac{\partial H_{\nu}(\Gamma(t))}{\partial\nu}\right|_{\nu=\nu(t)}\cdot
\frac{d\nu(t)}{dt}
\label{e:work-def}
\end{equation} 
to the system.
We denote by 
\begin{equation}
\Qt(\pathG)=\inttm J(\pathG;t)
\label{e:Qt}
\end{equation}
the heat transferred from the first to the second heat bath during $[-\ts,\ts]$.
Similarly, we write
\begin{equation}
\Qt^\mathrm{i}(\pathG)=\intti J(\pathG;t),
\quad
\Qt^\mathrm{f}(\pathG)=\inttf J(\pathG;t).
\label{e:Qt_i+f}
\end{equation}

The time evolution of the system is described by a Markov process.
We denote by $\WT_{\patha}[\pathG]$ the transition probability associated with
a path $\pathG$ in a protocol $\patha$.
It is normalized as $\int\calD\pathG\,\WT_{\patha}[\pathG]\delta(\Gamma(-\tl)-\Gi)=1$ 
for any initial state $\Gi$, 
where $\int\calD\pathG(\cdots)$ denotes the integral 
over all the possible paths $\pathG$.
For any function $f(\pathG)$, we define its average in the protocol $\patha$
as 
\begin{equation}
\sbkt{f}^{\patha}:=\int\calD\pathG\rhost_{\nu}(\Gamma(-\tl))\WT_{\patha}[\pathG]f(\pathG),
\end{equation}
where $\rhost_{\nu}(\Gamma)$ is the probability distribution for the unique NESS corresponding to the parameters $\nu=\nu(-\tl)$.

\subsection{Examples of the system}

To be concrete we describe two typical examples to which our  theory apply.
We stress however that the theory is quite general and depends only on the detailed fluctuation theorem.

\paragraph*{Heat conducting system with thermal walls:}
We first discuss the heat conducting system as shown in Fig.~\ref{fig:fig1}.
A rectangular box is filled with $N$ particles with an equal mass $m$. 
Two walls in contact with heat baths are designed as thermal walls 
and others as normal reflective walls.
When a particle collides with a thermal wall in contact with a heat bath
of $\beta_k$ ($k=1, 2$),
its velocity $\mathbf v$
is randomized according to the probability density
\begin{eqnarray}
f_k({\mathbf v})\propto
|v_{\perp}|\exp\left[-\frac{\beta_k m |{\mathbf v}|^2}{2}\right].
\end{eqnarray}
where $v_{\perp}$ is the component of $\mathbf v$ perpendicular to the thermal wall. We assume elastic collisions in the reflective walls.

The energy exchange between the system and the thermal wall with $\beta_k$ ($k=1,2$) is described as follows.
Suppose that we have $n_k$ collisions between time $t$ and $t+\Di t$,
and that the velocities of the particle changes from ${\mathbf v}_0^i$ 
to ${\mathbf v}^i$ in the $i$-th collision ($0\le i\le n_k$).
The heat current from the k-th heat bath to the system is then defined as
\begin{eqnarray}
J_k(\pathG;t) = \frac{1}{\Di t}\sum_{i=1}^{n_k} \frac{m(|{\mathbf v}^i|^2-|{\mathbf v}_0^i|^2)}{2},
\end{eqnarray}
where $\Di t$ is taken sufficiently small.

One of the particle indexed by $\mathrm{A}$ (green in Fig.~\ref{fig:fig1})
is under the potential for the external operation 
\begin{eqnarray}
V_{\nu}(\mathbf{r}_{\mathrm{A}})&=&\frac{k}{2}|\mathbf{r}_{\mathrm{A}}-\nu|^2
,
\end{eqnarray}
as well as the potential $V({\mathbf{r}_1,\cdots,\mathbf{r}_\mathrm{A},\cdots,\mathbf{r}_N})$
for the interaction with surrounding particles.
The parameter $\nu$ gives the equilibrium position for the particle A.
By operating the value of $\nu$, we can measure 
the mechanical work $W(\pathG)$.
The present stochastic model is often used in numerical simulations of heat conduction.  See, e.g., \cite{thermalWall}.
It is not difficult to show that the present model satisfies the detailed fluctuation theorem.  See, e.g., \cite{thermalWall-KNST}.

\paragraph*{Driven system:}
We next discuss a well-studied model other than heat conduction, 
i.e., the system of Brownian particle under controllable potential and 
constant driving field. 
Consider a particle in a 1-dimensional periodic potential $V_{\nu}(x)$
with a periodic boundary condition. 
The particle is driven by a constant external field $F$
besides the force from the potential $V_{\nu}$.
Its time evolution is governed by the Langevin equation,
\begin{eqnarray}
\gamma\dot x=-\frac{dV_{\nu}}{dx}+F+\sqrt{\frac{2\gamma}{\beta}}\xi(t),
\end{eqnarray}
where $\xi(t)$ is a Gaussian white noise with$\sbkt{\xi(t)}=0$ and $\sbkt{\xi(t)\xi(t')}=\delta(t-t')$.

In this system, we do not need to measure heat current. 
Instead, the important observable is the work performed by the constant field
\begin{eqnarray}
W_\mathrm{F}(\pathG):=\inttm  F \dot x.
\end{eqnarray}
Then all the results in the present paper, including the extended Jarzynski equality, are valid in the present system if one replaces 
$\mbeta$ with $\beta$, and 
$\Di\beta \Qt(\pathG)$ with $\beta W_\mathrm{F}(\pathG)$, respectively.  
By modifying the shape of the potential $V_{\nu}(x)$ externally 
via the change of the parameter $\nu$, 
mechanical work $W(\pathG)$ (given by \eqref{e:work-def})
is performed to the system.

\section{Jarzynski equality for NESS}

In \cite{KN-long},
we introduced a nonequilibrium free energy $F(\nu)$ 
which is a function of the parameter $\nu$ (as well as $\beta_1$ and $\beta_2$),
and coincides with the equilibrium free energy for $\beta_1=\beta_2$.
Here we show that, for any $\beta_1$, $\beta_2$, and operation $\patha$, the exact identity
\begin{eqnarray}
\bbkt{e^{-\mbeta(W-\Di F)}}^{\patha}=\bbkt{e^{\dbeta\,\Qt}}^{\pathad}_\mathrm{m}
,
\label{e:Jarzynski}
\end{eqnarray}
is valid, where  $\Di F :=F(\nu')-F(\nu)$.
The identity \eqref{e:Jarzynski} is our most basic result.
Here we introduced a modified expectation 
\begin{eqnarray}
\bbkt{f}^{\patha}_\mathrm{m}
:=
\frac{\bbkt{e^{\dbeta\,\Qt^\mathrm{i}/2} ~f~ e^{\dbeta\,\Qt^\mathrm{f}/2}}^{\patha}}{\bbkt{e^{\dbeta\,\Qt^\mathrm{i}/2} ~e^{\dbeta\,\Qt^\mathrm{f}/2}}^{\patha}}
,
\label{e:mExpectation}
\end{eqnarray}
where $f(\pathG)$ is an arbitrary function \cite{timeReversalSymmetry}.
See the Appendix for the derivation of Eq.~\eqref{e:Jarzynski}.

For equilibrium operations
with $\beta_1=\beta_2$ (i.e. $\dbeta=0$ and $\mbeta=\beta_1=\beta_2$),
it is obvious that 
\eqref{e:Jarzynski} reduces to the celebrated Jarzynski equality
\begin{equation}
\bbkt{e^{-\mbeta(W-\Di F)}}^{\patha}=1.
\end{equation}
We would like to propose \eqref{e:Jarzynski} as the most natural nonequilibrium extension of the Jarzynski equality to NESS.

Let us stress that $W(\pathG)$ is the standard mechanical work,
and the left-hand side of \eqref{e:Jarzynski} can be evaluated experimentally
(exactly as in the case of the original Jarzynski equality).
Although the right-hand side may appear artificial,
we shall show below that this quantity can also be evaluated experimentally
 (at least when the NESS is close to equilibrium).
To be specific, we show below in \eqref{e:cumulant} that this modified expectation
is directly connected to the well-known linear response relation 
by rewriting the modified expectation
in terms of heat currents $J(t)$.

We emphasize that the time-reversed protocol $\pathad$ is experimentally executable as well as the protocol $\pathad$.
The average $\sbkt{\cdots}^{\pathad}$ is measurable
in a physically natural time evolution with the time-reversed protocol.

We also note that since $W(\pathG)=0$ for a constant protocol $\patha=(\nu)$, 
the equality \eqref{e:Jarzynski} implies 
\begin{equation}
\sbkt{e^{\dbeta\,\Qt}}^{(\nu)}_\mathrm{m}=1.
\label{e:response}
\end{equation}
This is a version of the integrated fluctuation theorem 
and is related to a response relation as usual (see \cite{FTcomment}).
We can say that the equality \eqref{e:Jarzynski} reveals an intimate relation 
between the mechanical work in NESS and the response relation.
We shall see later that a deviation of $\sbkt{e^{\dbeta\,\Qt}}^{\patha}_\mathrm{m}$ from $1$ corresponds to a violation of the response relation.

\subsection{Quasi-static limit}

Before proceeding to the interpretation of the right hand side of \eqref{e:Jarzynski}, 
we demonstrate straightforward conclusions of the extended Jarzynski equality \eqref{e:Jarzynski}.

Since $W(\pathG)$ essentially does not fluctuate in the quasi-static limit,
one has $\sbkt{e^{-\mbeta W}}^{\patha}=e^{-\mbeta\sbkt{W}^{\patha}}$.
By also noting that $\sbkt{W}^{\pathad}=-\sbkt{W}^{\patha}$ in this limit,
\eqref{e:Jarzynski} reduces to
\begin{eqnarray}
\bbkt{W}^{\patha}
=
\Di F
+\mbeta^{-1}\log\bbkt{e^{\dbeta\,\Qt}}^{\patha}_\mathrm{m}
,
\label{e:Gibbs}
\end{eqnarray}
which is an exact relation corresponding to the Gibbs relation in equilibrium
thermodynamics.

The equilibrium Gibbs relation leads to potentials which describe macro- or mesoscopic forces.
The equality \eqref{e:Gibbs}, however, implies that 
we may not have such potentials in NESS as $\log\bbkt{e^{\dbeta\,\Qt}}^{\patha}_\mathrm{m}$ 
is not necessarily described by a difference of a state function.

\subsection{The second law for NESS}

From Jensen's inequality, we have 
$\log\sbkt{e^{-\mbeta(W-\Di F)}}^{\patha}\ge -\mbeta(\sbkt{W}^{\patha}-\Di F)$,
which implies
\begin{eqnarray}
\bbkt{W}^{\patha} \ge \Di F-\mbeta^{-1}\log\bbkt{e^{\dbeta\,\Qt}}^{\pathad}_\mathrm{m}
,
\label{e:2ndLaw}
\end{eqnarray}
where the equality holds in the quasi-static limit.
We believe that this is a natural extension of the second law of thermodynamics to operations between NESS.
It is notable that the right-hand side involves a quantity in the reversed protocol $\pathad$.

The inequality \eqref{e:2ndLaw} implies that
the minimum work principle is not extended straightforwardly to NESS.
The quantity equated with the free energy difference is not the work but
the sum of the work and $\mbeta^{-1}\log\sbkt{e^{\dbeta\,\Qt}}^{\pathad}_\mathrm{m}$.
This apparently means that one must invoke the reversed protocol $\pathad$
to find the limitation of the work.

\section{Expansion in weak nonequilibrium regime}

We define a dimensionless parameter indicating the degree of nonequilibrium
by 
\begin{equation}
\epsilon:=|\dbeta|/\mbeta.
\end{equation}
We here deal with systems with small $\epsilon$
and ignore the contribution of $\oep$.

Let us now derive a compact approximate expression \eqref{e:cumulant}
for the right-hand side of \eqref{e:Jarzynski}.
From the definition \eqref{e:mExpectation} of the modified expectation ,
we have
\begin{eqnarray}
\log\bbkt{e^{\dbeta\,\Qt}}^{\patha}_\mathrm{m}
=
\log\bbkt{e^{\dbeta\,\Qt^\mathrm{i}/2} e^{\dbeta\,\Qt} e^{\dbeta\,\Qt^\mathrm{f}/2}}^{\patha}
-
\log\bbkt{e^{\dbeta\,\Qt^\mathrm{i}/2} e^{\dbeta\,\Qt^\mathrm{f}/2}}^{\patha}
.
\end{eqnarray}
By applying the cumulant expansion to the right-hand side
and arranging the result by order,
we have 
\begin{eqnarray}
\log\bbkt{e^{\dbeta\,\Qt}}^{\patha}_\mathrm{m}
=
\dbeta\bbkt{\Qt}^{\patha}
+
\frac{\dbeta^2}{2}\bbkt{\Qt; (\Qt^\mathrm{i}+\Qt+\Qt^\mathrm{f})}^{\patha}
+\oep
,
\label{e:cumulant0}
\end{eqnarray}
where $\sbkt{A;B}$ is a truncated correlation, $\sbkt{A;B}:=\sbkt{AB}-\sbkt{A}\sbkt{B}$.
From eqs.~\eqref{e:Qt} and \eqref{e:Qt_i+f},
$\Qt(\pathG)=\inttm J(\pathG;t)$ and $\Qt^\mathrm{i}(\pathG)+\Qt(\pathG)+\Qt^\mathrm{f}(\pathG)=\intt J(\pathG;t)$.
Substituting them into eq.~\eqref{e:cumulant0}, we have
\begin{eqnarray}
\log\bbkt{e^{\dbeta\,\Qt}}^{\patha}_\mathrm{m}
=\dbeta\inttm \Jv^{\patha}(t) +\oep
,
\label{e:cumulant}
\end{eqnarray}
where we have defined
\begin{eqnarray}
\Jv^{\patha}(t)
&=&
\bbkt{J(t)}^{\patha} 
+\frac{\dbeta}{2} \int_{-\tl}^{\tl} ds \bbkt{J(t); J(s)}^{\patha}
,
\quad
(-\ts\le t \le \ts).
\label{e:LRviolation}
\end{eqnarray}

The equality \eqref{e:response} in the steady protocol $(\nu)$
is approximated as
\begin{equation}
\Jv^{(\nu)}(t)=0
\label{e:LRviolation-st}
\end{equation}
 (to be precise, $0$ should be read $\oet$),
which leads to the well-known formula of the linear response relation (LRR)
for heat currents \cite{Onsager,Kubo}.
Since $\tl\gg\ts+\tr$,
we have
\begin{eqnarray}
\frac{\dbeta}{2} \int_{-\tl}^{\tl} ds \bbkt{J(t); J(s)}^{(\nu)}
=\dbeta \int_{0}^{\infty} du \bbkt{J(t); J(t-u)}^{(\nu)}+\oet,
\end{eqnarray}
with which eq.~\eqref{e:LRviolation-st} is transformed to the 
usual Green-Kubo formula 
for the steady current $\sbkt{J(t)}^{(\nu)}$.
It is notable that
the modified expectation is regarded as a natural expectation
from the point of the LRR.

When there is an operation, 
the LRR is violated in general and  
$\Jv^{\patha}(t)$ does not vanish.
We can thus interpret $\Jv^{\patha}(t)$
as the ``{\it violation of LRR}'' due to the external operation.

More generally
the equality  $\sbkt{e^{\dbeta\,\Qt}}^{(\nu)}_\mathrm{m}=1$ in \eqref{e:response} 
gives an exact response relation for $\sbkt{J(t)}^{(\alpha)}$ 
because, by substituting eqs.~\eqref{e:Qt} and \eqref{e:Qt_i+f},
the equality \eqref{e:response} is rewritten as a relation which connects $\sbkt{J(t)}^{(\alpha)}$ to the higher cumulants of $J(\pathG;t)$.
Similarly to the above argument up to $\oet$, 
the deviation of $\sbkt{e^{\dbeta\,\Qt}}^{\patha}_\mathrm{m}$ from $1$
can be regarded as the violation of the exact response relation.
We have thus reached the most important interpretation of 
the equality \eqref{e:Jarzynski};
{\it the mechanical work in NESS is related to the violation of the response relation}.

In equilibrium operations at $\dbeta=0$, we see that $\Jv^{\patha}(t)=\sbkt{J(t)}^{\patha}$
corresponds to the heat current induced by the external operation.
This enables us to intuitively understand the role played by $\Jv^{\patha}(t)$ in a weak NESS.
In an equilibrium system the induced current $\sbkt{J(t)}^{\patha}$ requires no ``costs'', and hence does not appear in thermodynamic relations.
In a NESS, on the other hand, any heat current is coupled to the temperature difference.
This is the reason that we have $\dbeta\sbkt{J(t)}^{\patha}$ in our thermodynamic relation.

\section{Discussions}

We have derived nonequilibrium extensions of the Jarzynski work relation \eqref{e:Jarzynski},
the Gibbs relation \eqref{e:Gibbs}, and the second law \eqref{e:2ndLaw}
in a general classical model of heat conduction.
Although one can show that the Gibbs relation \eqref{e:Gibbs} approximated to $\oet$
coincides with the extended Clausius relation that we have derived in \cite{KNST-nl}, 
all the other results are novel.
Especially, it is fascinating that thermodynamic relations and response relations are coupled
intrinsically in our relations.
The traditional understanding has been that thermodynamic relations work for equilibrium operations, and 
response relations work for nonequilibrium steady states.
Here, by considering an operation for NESS and proving the extended Jarzynski
equality \eqref{e:Jarzynski}, 
we have shown that the two paradigms are naturally unified in a single exact relation.

There has been many works on the violation of fluctuation-dissipation relation including 
an effective temperature for characterizing
relaxation processes \cite{Cugliandolo-Kurchan},
the formula for estimating the energy dissipation \cite{Harada-Sasa,Speck-Seifert}, 
and 
the linear response around NESS \cite{Baiesi-Maes-Wynants,Prost-Joanny-Parrondo}.
It would be suggestive to look for possible relations of these topics
with the present study.

Although we have restricted ourselves to the simplest setting here,
it is straightforward to extend the present results to the case
where the inverse temperatures $\beta_1$ and $\beta_2$ vary, or 
to other nonequilibrium systems.
The only essential requirement is the detailed fluctuation theorem \eqref{e:symmetry}.

Last but not least let us stress that all of our main results 
\eqref{e:Jarzynski}, \eqref{e:Gibbs}, and \eqref{e:2ndLaw} may be tested
experimentally especially when the degree of nonequilibrium $\epsilon$ is small.
Although we still do not know whether these results have practical applications,
it would be exciting to imagine applying the exact Jarzynski equality \eqref{e:Jarzynski}
to the analysis of the efficiency of a thermodynamic machine operating in NESS.


The author thanks T. S. Komatsu, S. Sasa and H. Tasaki for fruitful discussions and suggestions.
This work was supported by Grants from the Ministry of Education, Science, Sports and Culture of Japan (19540392 and 23540435).

\section{Appendix}
\subsection{Derivation of the extended Jarzynski equality (12)}

We shall prove our main observation \eqref{e:Jarzynski}.
The proof relies on the detailed fluctuation theorem \eqref{e:symmetry} 
and the exact representation \eqref{e:KN} for the probability distribution of NESS.
It is essentially straightforward.

Let us decompose the time intervals as $[-\tl,\tl]=[-\tl,-\ts]\cup [-\ts,\ts] \cup [\ts,\tl]$, 
and, correspondingly, a path as $\pathG=(\pathGi,\pathGm,\pathGf)$.
We have 
$\Qt^\mathrm{i}(\pathG)=\Qt^\mathrm{i}(\pathGi)$ and $\Qt^\mathrm{f}(\pathG)=\Qt^\mathrm{f}(\pathGf)$
from eq.~\eqref{e:Qt_i+f}.
For an arbitrary function $f$ of $\pathGm=(\Gamma(t))_{t\in[-\ts,\ts]}$,
the numerator of the modified expectation \eqref{e:mExpectation} is naturally
decomposed as
\begin{eqnarray}
\bbkt{e^{\dbeta\Qt^\mathrm{i}(\pathG)/2}~f~ e^{\dbeta\Qt^\mathrm{f}(\pathG)/2}}^{\patha}
&=&
\int d\Gamma d\Xi~
\bbkt{e^{\dbeta\Qt^\mathrm{i}(\pathG)/2}}^{(\nu)}_{\ste,\Gamma}
\bsqbk{f}^{\patha}_{\Gamma,\Xi}
\bbkt{e^{\dbeta\Qt^\mathrm{f}(\pathG)/2}}^{(\nu')}_{\Xi,\ste}
,
\label{e:decompose}
\end{eqnarray}
by applying 
$\int d\Gamma d\Xi ~\delta(\Gamma(-\ts)-\Gamma)\delta(\Gamma(\ts)-\Xi)=1$
to the left hand side of \eqref{e:decompose}.
Here we defined conditioned expectations,
\begin{equation} 
\sbkt{f}_{\Gamma,\ste}^{(\nu)}
=\int \calD\pathGf\,\delta(\Gamma(\ts)-\Gamma)\WT_{(\nu)}[\pathGf]\,f(\pathGf), 
\end{equation}
with a fixed final state $\Gamma$ 
and
\begin{equation}
\sbkt{f}_{\ste,\Gamma}^{(\nu)}
=\{\rhost_{\nu}(\Gamma)\}^{-1}\int \calD\pathGi\,\rhost_{\nu}(\Gamma(-\tl))\,\WT_{(\nu)}[\pathGi]\delta(\Gamma(-\ts)-\Gamma)\,f(\pathGi),
\end{equation}
with an fixed initial state $\Xi$.
We also introduced unnormalized expectation $\sqbk{\cdots}$ by
\begin{eqnarray}
\bsqbk{f}^{\patha}_{\Gamma,\Xi} 
=
\rhost_{\nu}(\Gamma)\int \calD\pathGm f(\pathGm)\delta(\Gamma(-\ts)-\Gamma)
\delta(\Gamma(\ts)-\Xi) \WT_{\patha}[\pathGm]
,
\label{e:sqExpectation}
\end{eqnarray}
where $\WT_{\patha}[\pathGm]$ is the transition probability for $\pathGm$.

The transition probability $\WT_{\patha}[\pathGm]$ is known to satisfy the detailed fluctuation theorem,
\begin{eqnarray}
\WT_{\patha}[\pathGm]
~{e}^{\sum_{k=1}^2 \beta_k Q_k(\pathGm)}
=
\WT_{\pathad}[\pathGmd]
,
\label{e:symmetry}
\end{eqnarray}
where 
$\WT_{\pathad}[\pathGmd]$ is the transition probability of the time reversed path
$\pathGmd=(\Gamma(-t))_{t\in[-\ts,\ts]}$ for the reversed protocol $\pathad$.
$Q_k(\pathGm)=\inttm J_k(\pathGm;t)$ is the total heat 
that flows from the bath $k$ to the system during the path $\pathGm$.
By using the energy conservation
$H_{\nu'}(\Gamma(\ts))-H_{\nu}(\Gamma(-\ts))=
W(\pathGm)+\sum_{k=1}^2 Q_k(\pathGm)$,
\eqref{e:symmetry} is rewritten as
\begin{eqnarray}
e^{-\mbeta H_\nu(\Gamma(-\ts))-\mbeta W(\pathGm)}\WT_{\patha}[\pathGm]
=
e^{-\mbeta H_{\nu'}(\Gamma(\ts))+\dbeta \Qt(\pathGmd)}\WT_{\pathad}[\pathGmd]
,
\end{eqnarray}
where we noted that $\Qt(\pathGmd)=-\Qt(\pathGm)$.
By integrating over $\pathGm$ with constraints $\Gamma(-\ts)=\Gamma$
and $\Gamma(\ts)=\Xi$, 
\begin{eqnarray}
&&{e^{-\mbeta H_{\nu}(\Gamma)}}
\int \calD\pathGm 
e^{-\mbeta W(\pathGm)}
\delta(\Gamma(-\ts)-\Gamma)\delta(\Gamma(\ts)-\Xi) \WT_{\patha}[\pathGm]
\\
&=&
{e^{-\mbeta H_{\nu'}(\Xi)}}
\int \calD\pathGmd 
e^{\dbeta\Qt(\pathGmd)}
\delta(\Gamma(\ts)-\Gamma^*)\delta(\Gamma(-\ts)-\Xi^*) \WT_{\pathad}[\pathGmd].
\end{eqnarray}
Recalling \eqref{e:sqExpectation},
we have
\begin{eqnarray}
\frac{e^{-\mbeta H_{\nu}(\Gamma)}}{\rhost_{\nu}(\Gamma)}\bsqbk{e^{-\mbeta W}}^{\patha}_{\Gamma,\Xi}
=
\frac{e^{-\mbeta H_{\nu'}(\Xi)}}{\rhost_{\nu'}(\Xi^*)}\bsqbk{e^{\dbeta\Qt}}^{\pathad}_{\Xi^*,\Gamma^*}
.
\label{e:sqExpectation2}
\end{eqnarray}

In \cite{KN-long}, we derived (also by using the detailed fluctuation theorem \eqref{e:symmetry}) 
the exact representation for the probability distribution
in NESS with parameters $\nu$, 
\begin{eqnarray}
\rhost_{\nu}(\Gamma)
=e^{\mbeta (F(\nu)-H_{\nu}(\Gamma))}
\frac{\bbkt{e^{\dbeta\Qt^\mathrm{f}/2}}^{(\nu)}_{\Gamma^*,\ste}}
{\bbkt{e^{\dbeta\Qt^\mathrm{i}/2}}^{(\nu)}_{\ste,\Gamma}}
,
\label{e:KN}
\end{eqnarray}
where the normalization factor $F(\nu)$ was identified 
as a nonequilibrium free energy.
By substituting \eqref{e:KN} into \eqref{e:sqExpectation2},
one gets
\begin{eqnarray}
e^{\mbeta F(\nu)}
\bbkt{e^{\dbeta\Qt^\mathrm{i}/2}}^{(\nu)}_{\ste,\Gamma}
\bsqbk{e^{-\mbeta W}}^{\patha}_{\Gamma,\Xi}
\bbkt{e^{\dbeta\Qt^\mathrm{f}/2}}^{(\nu')}_{\Xi,\ste}
=
e^{\mbeta F(\nu')}
\bbkt{e^{\dbeta\Qt^\mathrm{i}/2}}^{(\nu')}_{\ste,\Xi^*}
\bsqbk{e^{\dbeta\Qt}}^{\pathad}_{\Xi^*,\Gamma^*}
\bbkt{e^{\dbeta\Qt^\mathrm{f}/2}}^{(\nu)}_{\Gamma^*,\ste}
\end{eqnarray}
By integrating over $\Gamma$ and $\Xi$, and using \eqref{e:sqExpectation},
this implies
\begin{eqnarray}
\bbkt{e^{\dbeta\Qt^\mathrm{i}/2} e^{-\mbeta (W-\Di F)}e^{\dbeta\Qt^\mathrm{f}/2}}^{\patha}
=
\bbkt{e^{\dbeta\Qt^\mathrm{i}/2} e^{\dbeta \Qt}e^{\dbeta\Qt^\mathrm{f}/2}}^{\pathad}
.
\end{eqnarray}
Since the operation takes place in the interval $[-\tob,\tob]$,
there is no correlation between $W(\pathG)$ and $e^{\dbeta\,\Qt^\mathrm{i}/2}$ or
$e^{\dbeta\,\Qt^\mathrm{f}/2}$
so that 
the right hand side is decomposed as 
\begin{eqnarray}
\bbkt{e^{\dbeta\Qt^\mathrm{i}/2} e^{-\mbeta (W-\Di F)}e^{\dbeta\Qt^\mathrm{f}/2}}^{\patha}
=\sbkt{e^{-\mbeta (W-\Di F)}}^{\patha}\sbkt{e^{\dbeta\Qt^\mathrm{i}/2} e^{\dbeta\Qt^\mathrm{f}/2}}^{\patha}.
\end{eqnarray}
Noting that $\sbkt{e^{\dbeta\Qt^\mathrm{i}/2} e^{\dbeta\Qt^\mathrm{f}/2}}^{\patha}
=\sbkt{e^{\dbeta\Qt^\mathrm{i}/2} e^{\dbeta\Qt^\mathrm{f}/2}}^{\pathad}$ \cite{separate},
we have
\begin{eqnarray}
\bbkt{e^{-\mbeta(W-\Di F)}}^{\patha}=
\frac{\bbkt{e^{\dbeta\Qt^\mathrm{i}/2} e^{\dbeta \Qt}e^{\dbeta\Qt^\mathrm{f}/2}}^{\pathad}}
{\bbkt{e^{\dbeta\Qt^\mathrm{i}/2} e^{\dbeta\Qt^\mathrm{f}/2}}^{\pathad}}
.
\label{e:Jarzynski-pre}
\end{eqnarray}
The right hand side of \eqref{e:Jarzynski-pre} can be replaced 
by $\sbkt{e^{\dbeta\,\Qt}}^{\pathad}_\mathrm{m}$
according to the definition \eqref{e:mExpectation} for the modified expectation. 
Then we have the desired \eqref{e:Jarzynski}.



\begin{thebibliography}{10}


\bibitem{Oono-Paniconi}
{Y. Oono and M. Paniconi},
{Prog. Theor. Phys. Suppl.}
\textbf{130}, 29 (1998).

\bibitem{Sasa-Tasaki}
S. Sasa and H. Tasaki,
J. Stat. Phys. {\bf 125}, 125 (2006).

\bibitem{Ruelle}
D. Ruelle,
Proc. Natl. Acad. Sci. U.S.A.  {\bf 100}, 3054 (2003).

\bibitem{Hatano-Sasa}
{T. Hatano and S. Sasa},
{Phys. Rev. Lett.}
\textbf{86}, 3463 
{(2001)}.



\bibitem{KN}
{T. S. Komatsu and N. Nakagawa},
Phys. Rev. Lett. {\bf 100}, 030601 (2008). arXiv:0708.3158 

\bibitem{KNST}
T. S. Komatsu, N. Nakagawa, S. Sasa and H. Tasaki,
Phys. Rev. Lett., {\bf 100}, 230602 (2008). arXiv:0711.0246



\bibitem{KNST-nl}
T. S. Komatsu, N. Nakagawa, S. Sasa and H. Tasaki,
J. Stat. Phys. {\bf 142}, 127 (2011).
arXiv:1009.0970


\bibitem{symmetry2}
D. J. Evans,  E. G. D. Cohen,  and G. P. Morriss,
{Phys. Rev. Lett.} 
\textbf{71},  2401 (1993);
G. Gallavotti and E. G. D. Cohen,
{ Phys. Rev. Lett.}
\textbf{74},   2694 (1995);
J. Kurchan,  
{J. Phys. A: Math. Gen.}
\textbf{31},   3719 (1998);
C. Maes, 
{J. Stat. Phys.}  
\textbf{95},  367 (1999);
C. Jarzynski, J. Stat. Phys. {\bf 98}, 77 (2000).





\bibitem{Jarzynski}
C. Jarzynski,
{ Phys. Rev. Lett.}
\textbf{78},   2690 (1997).

\bibitem{Crooks}
{G. E. Crooks},
{Phys. Rev. E}
\textbf{61},
{2361} {(2000)}.




\bibitem{e:J}
For the definition of heat current for the Langevin bath, see K. Sekimoto: 
{Stochastic Energetics}
{(Springer, Berlin, 2010)},
or {K. Sekimoto},
{J. Phys. Soc. Jpn.}
\textbf{66},
{1234}
{(1997)}
.

\bibitem{thermalWall}
T. Murakami, T. Shimada, S. Yukawa and N. Ito,
{J. Phys. Soc. Jpn.}
\textbf{72},
{1049} {(2003)}.

\bibitem{thermalWall-KNST}
T. S. Komatsu, N. Nakagawa, S. Sasa, H. Tasaki and N. Ito,
{Prog. Theor. Phys. Suppl.}
\textbf{184},
{329} {(2010)}



\bibitem{KN-long}
T. S. Komatsu, N. Nakagawa, S. Sasa and H. Tasaki,
J. Stat. Phys. {\bf 134}, 401 (2009).
arXiv:0805.3023

\bibitem{timeReversalSymmetry}
We remark that the extra factor $e^{\dbeta\,\Qt^\mathrm{i}/2}  e^{\dbeta\,\Qt^\mathrm{f}/2}$ 
has an effect of recovering the time-reversal symmetry 
and making the state ``as close as equilibrium'' in the relevant time intervals.

\bibitem{FTcomment}
Usually the fluctuation theorem is 
expressed as $P(+\Sigma)/P(-\Sigma)=e^{\Sigma}$,
where $P(\Sigma)$ is the probability that the entropy production in the baths
is $\Sigma$.
By integrating over $\Sigma$, we have $\sbkt{e^{-\Sigma}}=1$.
Since $-\dbeta \Qt$ is the entropy production due to the heat current,
$\sbkt{e^{\dbeta\Qt}}_\mathrm{m}^{(\nu)}=1$ can be regarded as a version of 
the integrated fluctuation theorem.




\bibitem{Kubo}
{R. Kubo, M. Toda and N. Hashitsume}:
{Statistical Physics II: Noneqilibrium Statical Mechanics}
{(Springer-Verlag, Berlin, 1991)}.

\bibitem{Onsager}
{L. Onsager},
{ Phys. Rev. }
\textbf{37},   405 (1931);
{ \it{ibid} }
\textbf{38},   2265 (1931).



\bibitem{Cugliandolo-Kurchan}
L. F. Cugliandolo and J. Kurchan,
J. Phys. A,
\textbf{27} (1994) 5749.
L. F. Cugliandolo, J. Kurchan and L. Peliti,
Phys.Rev.E,
\textbf{55} (1997) 3898.

\bibitem{Harada-Sasa}
T. Harada and S. Sasa,
Phys.Rev.Lett.,
\textbf{95} (2005)  130602.


\bibitem{Speck-Seifert}
T. Speck and U. Seifert,
Europhys. Lett.,
\textbf{74} (2006) 391.


\bibitem{Baiesi-Maes-Wynants}
M. Baiesi, C. Maes and B. Wynants,
Phys.Rev.Lett.,
\textbf{103} (2009) 010602.

\bibitem{Prost-Joanny-Parrondo}
J. Prost, J. F. Joanny and J. M. R. Parrondo,
Phys.Rev.Lett.,
\textbf{103} (2009) 090601.

\bibitem{separate}
Since $\nu(t)$ is constant except for $t\in [-\tob,\tob]$,
both the quantities are equal to 
$\sbkt{e^{\dbeta\Qt^{\mathrm{i}/2}}}^{(\nu)}\sbkt{e^{\dbeta\Qt^{\mathrm{f}/2}}}^{(\nu')}$.



\end{thebibliography}
\end{document}